\documentclass[preprint2]{aastex}

\usepackage{natbib}
\usepackage{graphicx}
\usepackage{amsmath}
\newcommand{\kmpsb}{km\,s$^{-1}$ }

\newcommand{\ndhpb}{N$_2$H$^+$ }
\newcommand{\hddpb}{H$_2$D$^+$ }
\newcommand{\ddhpb}{D$_2$H$^+$ }

\newcommand{\ccb}{cm$^{-3}$ }
\newcommand{\scb}{cm$^{-2}$ }

\newcommand{\nddpb}{N$_2$D$^+$ }

\newcommand{\pdix}[1]{$\times$ 10$^{#1}$}

\newcommand{\nddp}{N$_2$D$^+$}

\newcommand{\hddp}{H$_2$D$^+$}
\newcommand{\htp}{H$_3^+$}
\newcommand{\htpb}{H$_3^+$ }

\newcommand{\dtpb}{D$_3^+$ }
\newcommand{\dcop}{DCO$^+$}
\newcommand{\dcopb }{DCO$^+$ }

\newcommand{\cc}{cm$^{-3}$}
\newcommand{\sqc}{cm$^{-2}$}

\shorttitle{H$_2$ and the age of interstellar dark clouds}
\shortauthors{L. Pagani, E. Roueff, \& P. Lesaffre}

\begin{document}

\title{Ortho--H$_2$ and the age of interstellar dark clouds}

\author{Laurent Pagani}
\affil{LERMA, UMR8112 du CNRS, Observatoire de Paris, 61, Av. de l\arcmin Observatoire, 75014 Paris, France.}
\email{laurent.pagani@obspm.fr}
\and
\author{Evelyne Roueff}
\affil{LUTh, UMR8102 du CNRS, Observatoire de Paris,
5, place Jules Janssen,
92195 Meudon cedex, France}
\and
\author{Pierre Lesaffre}
\affil{LERMA, UMR8112 du CNRS, Ecole Normale Sup\'erieure, 24 rue Lhomond, 75231 Paris Cedex 05, France}

\begin{abstract}
Interstellar dark clouds are the sites of star formation. Their main component, dihydrogen exists under two states, ortho and para. H$_2$ is supposed to form in the ortho:para ratio (OPR) of 3:1 and to subsequently decay to almost pure para--H$_2$ (OPR $\leq$ 0.001). Only if the H$_2$ OPR is low enough, will deuteration enrichment, as observed in the cores of these clouds, be efficient. The second condition for strong deuteration enrichment is the local disappearance of CO, which freezes out onto grains in the core formation process. We show that this latter condition does not apply to DCO$^+$, which, therefore, should be present all over the cloud. We find that an OPR $\geq$ 0.1 is necessary to prevent DCO$^+$  {large-scale apparition}. We conclude that the inevitable decay of ortho--H$_2$ sets an upper limit of $\sim$6 million years to the age of starless molecular clouds under usual conditions.
\end{abstract}

\keywords{ISM: abundances --- ISM: clouds --- evolution --- ISM: molecules --- astrochemistry}

\section{Introduction}
The low mass star formation process is relatively well understood, however its first step details remain uncertain, especially regarding the question of time and age. Still much debated questions are: how long it takes for a cloud to form from the diffuse atomic medium, how long a molecular cloud lives, how long it takes to form a condensation that will evolve into a prestellar core that subsequently collapses to form a protostar. Today, much debate exists upon the lifetime of clouds, e.g. \citet{Hartmann:2001p1195} defending a short lifetime (1 to a few million years) while \citet{Tassis:2004p1196} and \citet{Mouschovias:2006p337} claim a typical age of 10 million years, but most arguments are either of limited statistical significance or model-dependent (unknown magnetic field strength, small scale turbulence rapid dissipation, etc.). Clues are needed and while the usual chemical age modeling is not satisfying (problem of unknown initial conditions), we present here a simple constraint based on basic chemistry: the absence of DCO$^+$ in dark cloud envelopes can only be explained if the clouds are young enough and we propose to determine an upper limit for this age which is as much independent from the initial conditions as possible.

\section{The abundance of CO and DCO$^+$, and the deuterium chemistry}
Dark clouds contain dust grains embedded in interstellar gas, itself mostly molecular as these clouds are self-shielded against our Galaxy's UV background. These dark clouds are cold \citep[typically 10 K, with cores as low as 7 K,][]{Pagani:2003p150,Pagani:2005p537}, and deep inside, where the density reaches a few 10$^4$ \cc, the heavy species freeze out onto grains to form ices. Both these ices and the gas are subject to a rich chemistry. Recently, strong deuteration (deuterium enhancement with respect to hydrogen carriers: \dcopb versus HCO$^+$, HDCO and D$_2$CO versus H$_2$CO, etc.) has been recognized as a useful tool to better understand both the chemistry itself and the star formation process \citep{Bergin:2007p95,Ceccarelli:2007p418}. 

H$_2$ is the main hydrogen reservoir in dark clouds, and similarly, HD is the main deuterium reservoir. Its relative abundance to H$_2$ is $\sim$3 $\times$ 10$^{-5}$ \citep{2006ASPC..348...47H,2006ApJ...647.1106L}. \citet{1976RvMP...48..513W} was the first to invoke chemical deuterium enrichment via the important reaction 
\begin{equation}
\rm H_3^+ + HD \rightarrow H_2D^+ + H_2 + \Delta E
\end{equation}

$\Delta$E = 232 K when all species (reactants and products) are in their ground state. In cold dark clouds, the forward reaction is strongly favorable, creating \hddp /\htpb ratios much larger than the original HD/H$_2$ ratio. If CO is abundant (X[CO] $\approx$ 1.5 $\times$ 10$^{-4}$, where X[] denotes the relative abundance of a species with respect to H$_2$), \htpb reacts 5 times more often with CO than with HD, and forms HCO$^+$, strongly decreasing the abundance of \hddp. 

\begin{figure}
\plotone{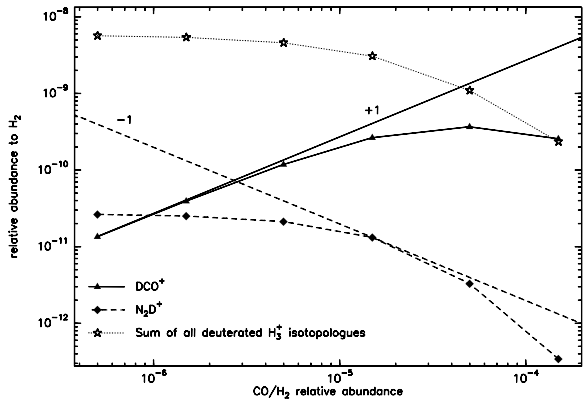}
\caption{DCO$^+$, N$_2$D$^+$ and H$_3^+$ isotopologues abundances as a function of CO abundance in a steady state model. Slopes of +1 and -1 are traced to better visualize the variation slope of the different ions. DCO$^+$, on the contrary to other deuterated species like N$_2$D$^+$ reaches its maximum abundance for high CO abundances.\label{fig1}}
\end{figure}

Once \hddpb is formed, it can be enriched in deuterium by reacting with HD again to form \ddhpb and \dtpb as first recognized by \citet{2003ApJ...591L..41R} :
\begin{eqnarray}
\rm H_2D^+ + HD \rightarrow D_2H^+ + H_2 + 187 K	\\
\rm D_2H^+ + HD \rightarrow D_3^+ + H_2 + 234 K
\end{eqnarray}
These two reactions are again favorable in the forward direction at low temperatures because of their exothermicity. Then, these \htpb  isotopologues can easily transfer their deuterons to other species:
\begin{eqnarray}
\rm CO + H_2D^+ \rightarrow \frac{2}{3}HCO^+  + \frac{1}{3}DCO^+ + \frac{1}{3}H_2 + \frac{2}{3}HD\\
\rm CO + D_2H^+  \rightarrow  \frac{1}{3}HCO^+ +  \frac{2}{3}DCO^+ +  \frac{1}{3}HD +  \frac{2}{3}D_2	\\
\rm CO + D_3^+ \rightarrow  DCO^+ + D_2	\\
\rm N_2 + H_2D^+ \rightarrow \frac{2}{3}N_2H^+  + \frac{1}{3}N_2D^+ + \frac{1}{3}H_2 + \frac{2}{3}HD\\
\rm N_2 + D_2H^+  \rightarrow  \frac{1}{3}N_2H^+ +  \frac{2}{3}N_2D^+ +  \frac{1}{3}HD +  \frac{2}{3}D_2	\\
\rm N_2 + D_3^+ \rightarrow  N_2D^+ + D_2
\end{eqnarray}
When CO is abundant, \hddpb is much less abundant as noted above and also reacts with CO preferentially, quenching all the other deuteration paths. CO also reacts with species like \ndhpb  and \nddpb to destroy them:
\begin{eqnarray}
\rm N_2H^+  + CO \rightarrow HCO^+  + N_2\\
\rm N_2D^+  + CO \rightarrow DCO^+  + N_2
\end{eqnarray}
The strong correlation between CO freeze-out onto grains and deuteration \citep[e.g.][]{2005ApJ...619..379C} confirm the chemical models \citep[e.g.][]{2003ApJ...591L..41R} as sketched above and has led to the conclusion that CO freeze-out was necessary for strong deuteration to occur. There is however one exception to this statement: the deuteration of CO itself can form \dcop. In first approximation, if \htpb is mainly destroyed by CO, X[\hddp ] varies like X[CO]$^{-1}$, whereas the production of \dcopb is proportional to X[CO] and X[\hddp ] (reaction 4). Thus the dependency upon X[CO] cancels out and X[\dcop] should remain approximately constant. Figure \ref{fig1} shows the abundance of X[\dcop], X[\htpb isotopologues] and X[\nddp] as a function of X[CO] in a steady-state chemical model. The X[\htpb isotopologues] is the sum of the abundances of all 3 deuterated \htpb isotopologues, which represent altogether the main deuteration partners of species like CO and N$_2$. The model is the same as in \citet{2008arXiv0810.1861P} but CO now runs from an undepleted relative abundance of 1.5 $\times$ 10$^{-4}$ down to a depletion by a factor 300 (X[CO] = 5 $\times$ 10$^{-7}$), for a cloud density of 10$^4$ \ccb at 10 K. When X[CO] starts to drop, X[\htpb isotopologues] increases faster than proportional (indicated by the -1 slope) because the \ddhpb abundance depends both on the destruction of \htpb by CO and on the destruction of \hddpb by CO, and the \dtpb abundance depends also on the destruction of \ddhpb by CO. This explains that at the beginning, X[\dcop] increases: the deuteration capabilities increase faster than X[CO] decreases. While X[CO] keeps decreasing, the deuteration capabilities reach saturation and X[\dcop] becomes proportional to X[CO] (traced by the +1 slope). X[\nddp], a representative gas--phase deuterated species, follows the deuteration capabilities of the cloud, as expected. It is clear that \dcopb has a behavior different from the other deuterated species, with an abundance varying only by ±50\% while the CO abundance drops by a factor of 30, and most importantly by \emph{ reaching its maximum abundance when CO is high} and deuteration capabilities are down. Therefore, one would expect \dcopb to be present everywhere in the cloud. It is not the case. Indeed, \dcopb is, as the other deuterated species, only detected in the depleted prestellar cores on extents similar to NH$_3$ \citep{1995ApJ...448..207B,2002A&A...382..583J} and nowhere else. This can only be explained if a high H$_2$ OPR is present. Indeed the H$_2$ OPR is an important controller of the cold cloud deuteration chemistry, as already discussed in \citet{1991MNRAS.248..173P,Flower:2006p82} and \citet{1992A&A...258..479P,2008arXiv0810.1861P}. 

\section{The role of ortho--H$_2$}

H$_2$ has 2 different possible nuclear spin states due to the proton spin of $\frac{1}{2}$ : parallel (ortho, I = 1, weight 2I+1 = 3 ) and anti-parallel (para, I = 0, weight = 1). The ortho state corresponds to the odd rotational levels (J = 1, 3, 5, É), the para state to the even levels. The first ortho state (J = 1) is 170 K above the ground (J = 0) para state. The ortho--para exchange is possible neither via radiative processes, nor via inelastic collisions. It can occur in the gas
phase and on the surface of dust grains \citep{2000A&A...360..656L}. The conversion rate on a solid surface is however highly uncertain
\citep{2010ApJ...714L.233W,2011PCCP...13.2172C} and it is most probable that this is mainly realized by gas-phase chemical reactions with H$^+$ or \htpb ions \citep{1973ApL....14...77D,1984MNRAS.209...25F,1991A&A...242..235L, Flower:2006p82}. The importance of ortho--H$_2$ in the deuteration process was first studied by \citet{1991MNRAS.248..173P,1992A&A...258..479P}, and later on, \citet{Flower:2006p82} revealed its importance as a deuteration regulator. However, they considered the consequences of the OPR evolution only for the prestellar core formation itself. In a study of L183 main prestellar core, \citet{2008arXiv0810.1861P}  noted that the H$_2$ OPR evolution timescale required by the observed \nddp/\ndhpb  ratio was on the order of the free-fall time. They also noted that the OPR had to drop below 0.01 for \nddpb to become detectable. 

The H$_2$ OPR value is not easily accessible. H$_2$ does not emit in cold dark clouds and is seen only in shocked regions, especially protostar outflows, or in absorption in UV in low extinction regions. The energy release from the H$_2$ formation reaction, assumed to be 1.5 eV in the equipartition hypothesis, is large enough to populate many levels of both ortho and para species: it is hence generally accepted, though not certain, that the OPR is 3:1 when H$_2$ forms on grains. Once released in the gas phase, H$_2$ tends to relax towards its thermal equilibrium state but never reaches it because at trace abundance levels, ortho--H$_2$ destruction is compensated by fresh H$_2$ formed onto grains from residual H and from destruction of ions such as \htpb and \hddpb that can release an ortho--H$_2$ \citep{1991A&A...242..235L, Flower:2006p82}. It is therefore necessary to model the chemical evolution of the H$_2$ OPR to understand the evolution of dark clouds. 
Once H$_2$ is formed, it can react with H$^+$ and \htpb to exchange states \citep{1991A&A...242..235L, Flower:2006p82}. For example,
\begin{equation}
\rm o\textendash H_2 + H^+ \rightarrow p\textendash H_2 + H^+ + \Delta E \\
\end{equation}
$\Delta$E = 170 K is the energy released as H$_2$ goes from J=1 to J=0 levels. The reverse endothermic reaction is thus difficult to obtain in a cloud at 10 K and ortho--H$_2$ slowly decreases to low abundance levels ($\approx$10$^{-3}$ -- 10$^{-4}$). Similarly, in reaction 1, except for HD, all the species have ortho and para states and $\Delta$E therefore depends in fact on the considered initial and final states. If only para states are involved in the products, $\Delta$E is maximal ($\Delta$E = 232 K with para--\htpb and 265 K with ortho--\htp, HD being always in its ground state, J = 0) but if both products are in their ortho ground state while the reactants are in their lowest state, para--\htpb ground state and HD (J = 0), $\Delta$E becomes negative and the reaction is slightly endothermic (\hddpb lowest ortho state is 87 K above the para ground state, and the total required energy is 25 K). Similarly, while the reverse reaction is impossible in a 10 K cloud with both para reactants, it becomes rapid with ortho--\hddpb and ortho--H$_2$ even at 10K, as this channel is slightly exothermic ($\Delta$E = 25 K) to produce para--\htp. A few such channels are opened with ortho--H$_2$ allowing the efficient destruction of the \htpb deuterated isotopologues. Ortho--H$_2$ is thus an important chemical poison to deuteration in dark clouds. Its abundance is therefore critical to control the deuteration enhancement \citep{Flower:2006p82,2008arXiv0810.1861P}. Because \dcopb is not observed everywhere in the clouds outside the depleted cores where it is detected \citep{1995ApJ...448..207B,2002A&A...382..583J}  while its abundance should be maximal, this sets a minimum level for the abundance of ortho--H$_2$. In turn, this minimum level gives rise to an upper limit to the age of the cloud, as ortho--H$_2$ abundance must eventually decrease to low levels as implied by reaction (12), and in order to explain the deuteration enhancements seen in the depleted cores, which, in some cold cores, can become huge \citep[amplification up to 10$^{12}$,][]{2002ApJ...571L..55L,2002A&A...388L..53V,2004A&A...416..159P}.

\section{The chemical model}
To address these questions, we used the deuteration network from \citet{2005A&A...438..585R,2007A&A...464..245R} in which we have included the ortho-para chemistry as described in \citet{2008arXiv0810.1861P}, based itself on the work by \citet{Walmsley:2004p558,Flower:2006p82} and the new rate coefficients of \citet{Hugo:2009p769}. We have also included analytical approximations of the spin state-dependent dissociative recombination rates of \htpb isotopologues from the tables in  \citet{2008arXiv0810.1861P}. The deuteration network includes all simple species (up to 6 atoms) based on C, N, O, and S with up to 5 deuterium substitutions (CD$_5^+$). For those reactions with a small endothermicity (E/k$_B <$ 600 K), we have corrected the endothermicity when the reaction with ortho--H$_2$ was involved (e.g., CD + H$_2$ $\rightarrow$ CH + HD is endothermic with para--H$_2$ and exothermic with ortho--H$_2$). For all reactions with rare species producing H$_2$ (e.g., HCN$^+$ + H $\rightarrow$ CN$^+$ + H$_2$), we have always assumed that H$_2$ is released in its para form, the most favorable energetically (similarly, only ortho--D$_2$ is considered in similar reactions with deuterated species). This does not noticeably accelerate the decay of ortho--H$_2$. We checked that after, e.g., 10$^5$ years, the para--H$_2$ production rate is 99.9\% dominated by reactions of ortho--H$_2$ with \htpb and H$^+$. Our model is not aimed at reproducing accurately a given cloud but it allows to follow the evolution of the H$_2$ OPR {in a molecule rich gas}. We have computed the chemical evolution of a cloud of constant density (n(H$_2$) = 10$^4$\,\cc), constant temperature (10\,K), average grain radius of 0.1\,$\mu$m and cosmic ionization rate of 1\,$\times$\,10$^{-17}$\,s$^{-1}$, which represent the usual conditions met in starless dark clouds. We have started with all species being atomic except H which is considered to be already entirely converted to H$_2$ and D to HD and we have considered different OPR starting ratios from 3 down to 3\,$\times$\,10$^{-3}$ to account for the possibility of H$_2$ production OPR below 3. Figure 2 shows the evolution of \dcopb and ortho--H$_2$ with time. We have traced the detectability limit of \dcopb for a cloud with a H$_2$ column density of 10$^{22}$\,\sqc, considering that a \dcopb J:1$\rightarrow$0 line with a width of 0.5\,\kmpsb and an intensity of 0.1\,K would be easily detectable with present day radiotelescopes. With the RADEX non-LTE radiative transfer model \citep{2007A&A...468..627V}, we find that a \dcopb column density of $\sim$2\,$\times$\,10$^{11}$\,\scb is detectable, equivalent to a relative abundance to H$_2$ of $\sim$2\,$\times$\,10$^{-11}$. Figure 2 shows that for any initial H$_2$ OPR below 0.1, \dcopb should be detected throughout the cloud in less than 3 \pdix{5} years. For an initial OPR = 0.1, the \dcopb becomes marginally detectable around $\sim$5\,$\times$\,10$^5$ years, disappears and finally becomes clearly detectable after 2\,My. For OPR $>$ 0.1, it takes 3 to 6 million years to become detectable, i.e. when the H$_2$ OPR drops below $\sim$0.03. 
Compared to the H$_2$ OPR drop from 3 to $\sim$0.01 in depleted cores which takes less than 2\,$\times$\,10$^5$\,years to happen \citep{2008arXiv0810.1861P}, the difference is twofold: as density increases during the prestellar core formation, 1) the chemistry accelerates, and 2) depletion sets in and H$^+$ is no more destroyed by the other species which have now disappeared from the gas phase (\htpb becomes also more abundant but to a lesser extent). H$^+$ can even become the dominant ion. The higher abundance of H$^+$ and \htpb and the accelerated chemistry provoke the rapid decline of ortho--H$_2$. 

On the same figure, we also trace the abundance of the sum of the deuterated \htpb isotopologues in the case H$_2$ OPR = 3. It shows that under 1 My, deuteration is dominated by the carbon chemistry (CH$_2$D$^+$, C$_2$HD$^+$, etc.) which is not dependent upon the ortho--H$_2$ abundance due to exothermicities close to 400 K. It is only beyond 1 My that the \htpb deuteration network takes over when ortho--H$_2$ starts to be negligible. If ortho--H2 is low since the beginning, then the \htpb isotopologue contribution to the abundance of DCO$^+$ is important from the start which explains the disappearance of the dip between 1 and $\sim$3 My.

The model is also sensitive to two other initial conditions:  the abundance of metals and the cosmic ray ionization rate. Metals, especially alkali and alkaline earth metals  like Ca, Na, K, and others like Fe, have a direct influence on the electronic equilibrium of the chemical model which in turn changes the dissociative recombination efficiency for DCO$^+$ and therefore its abundance. For standard $\zeta$ (1 \pdix{-17} s$^{-1}$) and H$_2$ OPR (3), we have varied their abundance from 3.4 \pdix{-8} to 1.3 \pdix{-7}. 
For the lowest abundance (50\% of the default one), DCO$^+$ becomes detectable much earlier (~2 \pdix{5} y) than in the standard case, drops below the detectability limit after ~1\,My and is detectable again after 4\,My. For the highest abundance (twice the default one), DCO$^+$ is detectable after 7.5 My. The cosmic ray ionization rate (Fig. \ref{fig3}): if the rate increases, the production of H$^+$ and \htpb increases, therefore accelerating the decay of ortho--H$_2$ and increasing the abundance of HCO$^+$ and DCO$^+$. For $\zeta$ = 3 \pdix{-17} s$^{-1}$, DCO$^+$ becomes detectable in less than 10$^5$ years. {Values of $\zeta$ in between 1 and 3 \pdix{-17} s$^{-1}$ would bring our model in agreement with some recent estimates of cloud lifetime \citep{Enoch:2008p316,2008A&A...482..855H}}. Conversely, for $\zeta$ = 3 \pdix{-18} s$^{-1}$, the chemistry is considerably slowed down, DCO$^+$ becomes detectable only after 28 My and steady-state is reached only after 47 My. It must be noted however that such a low ionization rate seems unlikely following \citet{Padovani:2009p557}.

\section{Conclusions}
Ortho--H$_2$ abundance is most probably the main limiting parameter for \dcopb production. The absence of \dcopb outside depleted cores, while its abundance is expected to be maximum for CO abundances close to 10$^{-4}$, is due to a high ortho--H$_2$ abundance in dark clouds (OPR $\geq$ 0.1) maintained long enough. Such high abundance can survive thanks to low H$^+$ and \htpb abundances, remaining low via charge or proton exchanges with many neutral species. However, ortho--H$_2$ eventually disappears. This sets an upper limit to the age of \dcop-quiet dark clouds of 3--6 million years after H$_2$ has formed for the most probable initial conditions. Higher cosmic ray ionization rates would increase the abundance of H$^+$ and \htpb and can thus only shorten the maximal possible age of the clouds to values which are too small while lower rates would tend to relax the age constraint  but seem very unlikely. This result strengthens the results of \citet{Hartmann:2001p1195,Enoch:2008p316,2008A&A...482..855H} and is not compatible with those of  \citet{Tassis:2004p1196} and \citet{Mouschovias:2006p337}. From the chemical point of view, cold dark clouds seem to be really short-lived.

\acknowledgments

We are grateful to Patrick Hennebelle for fruitful discussions. ER acknowledges the support of the ANR N0. 09-BLAN-020901.

\bibliographystyle{apj}

\clearpage

\begin{figure}
\epsscale{0.80}
\plotone{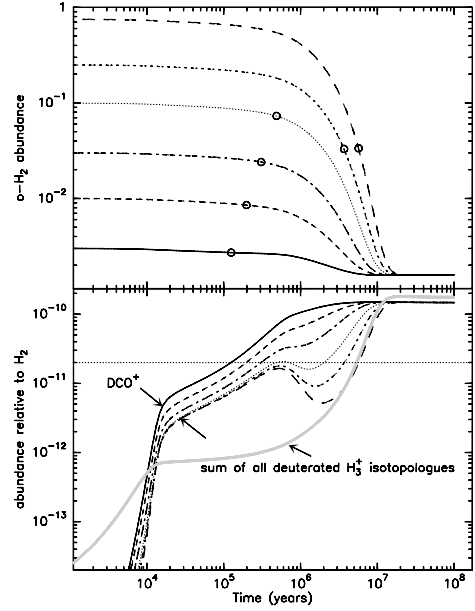}
\caption{Variation of DCO$^+$ and ortho---H$_2$ abundances in a pseudo time-dependent chemical model for different starting H$_2$ OPRs from 3 $\times$ 10$^{-3}$ up to 3. The abundance of the sum of the deuterated \htpb isotopologues is also traced as a thick gray line (for the case H$_2$ OPR = 3). The horizontal dotted line in the lower box indicates the limit of detection of DCO$^+$ and the circles in the upper box mark the corresponding OPR values.\label{fig2}}
\end{figure}

\begin{figure}
\plotone{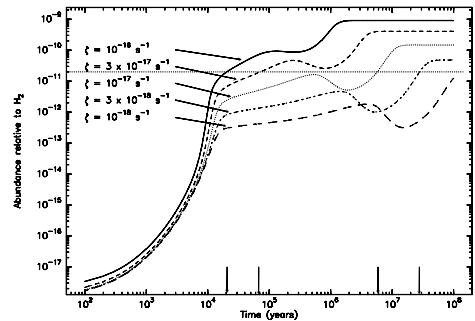}
\caption{Variation of DCO$^+$ and ortho---H$_2$ abundances in a pseudo time-dependent chemical model for different cosmic ray ionization rates from 1 $\times$ 10$^{-18}$ up to 1 \pdix{-16}. The horizontal dotted line indicates the limit of detection of DCO$^+$, the arrows mark the time at which DCO$^+$ abundance crosses the detection limit.\label{fig3}}
\end{figure}

\end{document}